\RequirePackage{fix-cm}
\documentclass[pdflatex,sn-mathphys]{sn-jnl}

\usepackage{graphicx}
\usepackage{float}
\usepackage[cmex10]{amsmath}
\usepackage{booktabs}
\usepackage{placeins}
\usepackage{xcolor}
\usepackage[T1]{fontenc}
\usepackage[nomath]{lmodern}
\ifdefined\XeTeXversion
  \DeclareFontShape{OML}{cmm}{m}{it}{<->cmmi10}{}
  \DeclareFontShape{OMS}{cmsy}{m}{n}{<->cmsy10}{}
  \DeclareFontShape{OT1}{cmr}{m}{n}{<->cmr10}{}
  \DeclareFontShape{OT1}{cmr}{bx}{n}{<->cmbx10}{}
\fi
\IfFileExists{manyfoot.sty}{\usepackage{manyfoot}}{}
\ifdefined\DeclareNewFootnote\else
  \newinsert\footinsA
  \newcommand{\SetFootnoteHook}[1]{}
  \NewDocumentCommand{\DeclareNewFootnote}{m o}{}
  \let\footnoteA\footnote
\fi

\providecommand{\ket}[1]{\lvert #1 \rangle}
\providecommand{\bra}[1]{\langle #1 \rvert}
\newcommand{\Hp}{H_{\mathrm{p}}}

\newcommand{\MI}{\hat{I}}

\newcommand{\best}[1]{{\boldmath$#1$}}

\begin{document}

\title[Selecting clinical features with measurement-guided quantum feedback]{Selecting clinical features with measurement-guided quantum feedback}

\author*[1]{\fnm{Pedro M.} \sur{Prado}}\email{pprado@estudante.ufscar.br}
\equalcont{These authors contributed equally to this work.}

\author*[2]{\fnm{Giovanni} \spfx{dos} \sur{Santos Franco}}\email{gs.franco@unesp.br}
\equalcont{These authors contributed equally to this work.}

\author[2,3]{\fnm{Felipe F.} \sur{Fanchini}}

\author[1]{\fnm{Celso J.} \sur{Villas-Boas}}\email{celsovillasboas@ufscar.br}

\affil*[1]{\orgdiv{Departamento de Computa\c{c}\~ao}, \orgname{Universidade Federal de S\~ao Carlos}, \orgaddress{\postcode{13565-905}, \city{S\~ao Carlos}, \state{SP}, \country{Brazil}}}

\affil[2]{\orgdiv{School of Sciences}, \orgname{S\~ao Paulo State University}, \orgaddress{\postcode{17033-360}, \city{Bauru}, \state{SP}, \country{Brazil}}}

\affil[3]{\orgname{Hospital Israelita Albert Einstein}, \orgaddress{\city{S\~ao Paulo}, \state{SP}, \country{Brazil}}}

\abstract{In this work, we study quantum optimization algorithms applied to feature selection in clinical data, that is, to the choice of a small set of variables that is both informative and minimally redundant. In particular, we use non-variational, feedback-based algorithms. In these algorithms, measurements performed on the quantum state directly determine the parameters of each new circuit layer, without the need for a classical optimizer. To keep the circuits shallow, we combine this strategy with a measurement-guided initialization: the outcomes of each round are used to prepare the initial state of the next round. In addition, we investigate whether controls inspired by counterdiabatic driving can increase the probability of finding the optimal solution with few layers. As case studies, we consider two healthcare problems. The first is the identification of suspect or pathological tracings in cardiotocographic fetal monitoring. The second is distinguishing participants with autism spectrum disorder from neurotypical participants using metabolic markers. On both reference instances, adding a counterdiabatic-inspired two-body control to the transverse field increased the probability of sampling the optimal set of variables. The selected panels achieved predictive performance comparable to that obtained with exact classical optimization.}

\keywords{Feature selection, Quantum optimization, FALQON, Measurement-guided initialization, Clinical data}

\maketitle

\section{Introduction}\label{sec1}

Clinical prediction models frequently draw on several measurements derived from the same physiological signal or biochemical pathway. This work considers two such settings. In fetal monitoring, a cardiotocogram (CTG) is summarized by 21 variables describing fetal heart-rate accelerations, decelerations, baseline and variability, together with uterine contractions, and many of these variables are correlated by construction \citep{ctg}. In autism research, a metabolomic panel of folate-dependent one-carbon metabolism and transsulfuration (FOCM/TS) reports 22 biochemical measurements, including metabolite concentrations alongside ratios computed from them \citep{howsmon2017}. In both cases, a compact panel can simplify interpretation and, when its variables require distinct assays, reduce the number of measurements needed \citep{guyon2003,saeys2007}. Ranking variables one at a time by their association with the outcome, however, tends to select measurements that carry largely the same information.

Relevance--redundancy filters address this problem directly. The minimum-redundancy maximum-relevance (mRMR) principle \citep{peng2005} scores a subset by the mutual information between each selected feature and the label, penalized by the mutual information shared among the selected features. Because the redundancy penalty is pairwise, finding the best subset is a quadratic unconstrained binary optimization (QUBO) problem, a class that is NP-hard in general \citep{glover2019}. In practice, mRMR is optimized greedily, one feature at a time, and the gap between the greedy subset and the optimal one is rarely reported. For a clinical panel, moreover, the optimal subset and a good classifier are not the same thing. Two four-feature panels can have nearly the same relevance--redundancy score while exchanging a variable that matters to the classifier, so both the quality of the optimization and the predictive value of its output need to be measured.

A QUBO can be written as an Ising Hamiltonian whose ground states encode the minimizing bitstrings \citep{lucas2014,glover2019}. One gate-based approach to this problem uses variational quantum algorithms (VQAs). In the quantum approximate optimization algorithm (QAOA), a classical optimizer updates the circuit rotation angles to minimize the expected cost \citep{farhi2014,cerezo2021}. Depending on the circuit and objective, the search can also encounter small gradients or difficult optimization landscapes \citep{mcclean2018,bittel2021}.

Feedback-based quantum algorithms (FQAs) determine the controls differently. In the Feedback-based Algorithm for Quantum Optimization (FALQON), a measurement on the state prepared so far determines how to drive the next layer \citep{magann2022prl,magann2022pra}. The measured observable gives the contribution of a unit-amplitude driver to the instantaneous rate of change of the problem energy. Applying a control with the opposite sign gives a direction of energy descent in the continuous-time dynamics. Once the control law and its settings are fixed, this measurement specifies the new amplitude directly. There is no need to test competing angles with a classical optimizer. The cost shifts to constructing a sequence of layers and estimating feedback for successive circuit depths. On weighted MaxCut instances, standard FALQON required hundreds of layers in some runs \citep{rattighieri2026}.

\citet{rattighieri2026} introduced measurement-guided initialization (MGI) to carry information from one shallow feedback circuit to the next. A shallow circuit is executed for several rounds. Outcomes retained from one round determine the single-qubit biases of the product state prepared for the next. Information can therefore accumulate across rounds while the longest coherent evolution stays at a chosen depth. A second way to modify the feedback dynamics draws on counterdiabatic (CD) driving, which uses auxiliary Hamiltonian terms to suppress transitions during an adiabatic interpolation \citep{demirplak2003,berry2009,sels2017,claeys2019}. Related constructions have been studied in digitized optimization, QAOA and feedback protocols \citep{hegade2021,chandarana2022,cadavid2025,malla2024}.

Quantum optimization has been applied to DNA assembly \citep{boev2021,sarkar2021,prado2026} and lattice protein folding \citep{robert2021}. For feature selection, \citet{mucke2023} used gate-based and annealing hardware to solve a relevance--redundancy QUBO, tuning the trade-off between the two terms to select a prescribed number of features. \citet{hellstern2024} found that performance relative to classical solvers depended on the dataset. In simulations of neutral-atom arrays, \citet{orquin2026} encoded relevance through local detunings and redundancy through interatomic interactions, then selected fixed-size subsets from the measured bitstrings. Such comparisons depend on the evaluation protocol and the classical baselines \citep{gupta2025,schuld2022,bowles2024}.

Biomedical applications include radiomic feature selection \citep{felefly2023} and pixel selection in medical images \citep{nau2025}. Closest to the present setting, \citet{liu2026} examined fixed-size feature selection in three autism-related metabolomics datasets, comparing BF-DCQO and QAOA on gate-based hardware with exact search and classical selectors.

Here we apply MGI to clinical feature selection and ask how the choice of feedback driver affects both recovery of the optimal panel and its predictive value. Building on earlier comparisons of mixers for portfolio QUBOs \citep{prado2026qip}, we compare a transverse field with three extensions that add a uniform $Y$ field, a field-weighted $Y$ control or a control that also uses the pairwise Ising couplings. We enumerate all candidate panels to identify the QUBO optimum and measure how often each solver recovers it. How much probability does a five-layer circuit assign to the optimal panel after repeated initialization? Does the driver ranking persist when the training data or the available depth changes? Finally, when the panel is selected using training data alone, does greater accuracy in solving the QUBO improve discrimination on held-out observations?

The remainder of the paper is organized as follows. Section~\ref{sec2} formulates the relevance--redundancy QUBO and describes the feedback protocol, the CD-inspired drivers and MGI. Section~\ref{sec3} presents the datasets, simulations and evaluation protocol. Section~\ref{sec4} reports solver diagnostics and predictive results, Section~\ref{sec5} discusses their implications and limitations, and Section~\ref{sec6} concludes.

\section{Problem formulation and algorithms}\label{sec2}
\suppressfloats[t]

\begin{figure}[!htbp]
\centering
\includegraphics[width=\linewidth]{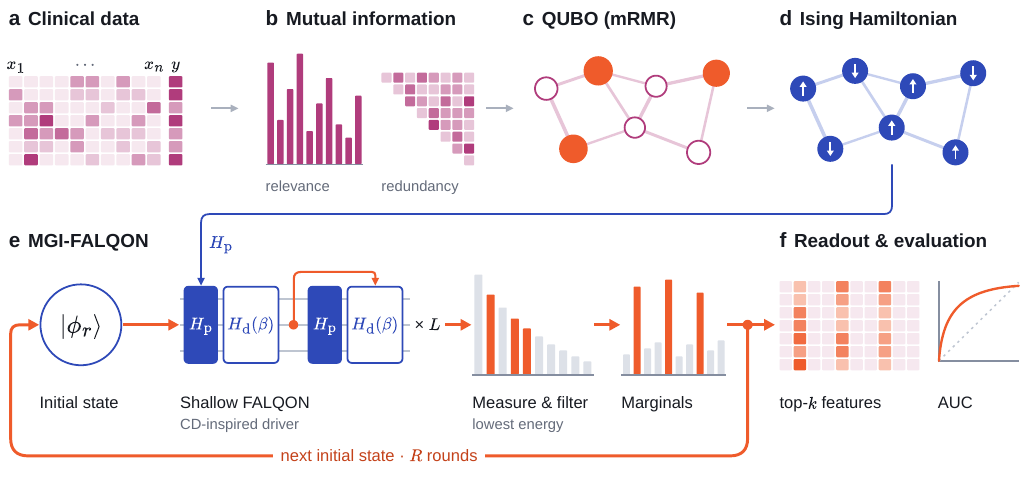}
\caption{Clinical feature selection from raw measurements to held-out evaluation. (a,b) The clinical records supply the feature values and class labels. Mutual information measures each feature's relevance to the label and its redundancy with other features. (c,d) These scores define a QUBO and its equivalent Ising Hamiltonian, with one qubit per feature. (e) After each shallow feedback circuit, low-energy candidates among the most probable bitstrings determine the feature inclusion probabilities used to initialize the next round. (f) After the final round, the four features with the largest filtered inclusion probabilities form the panel evaluated by logistic regression on held-out records.}\label{fig1}
\end{figure}

This section builds the optimization problem and the quantum protocol used to solve it. Section~\ref{subsec2.1} writes relevance--redundancy feature selection as a QUBO, and Section~\ref{subsec2.2} maps it onto an Ising Hamiltonian whose ground state encodes the optimal panel. Sections~\ref{subsec2.3}--\ref{subsec2.5} describe how FALQON, the CD-inspired drivers and MGI search for that ground state. Figure~\ref{fig1} summarizes the complete pipeline.

\FloatBarrier

\subsection{Relevance--redundancy QUBO}\label{subsec2.1}

Let $f_1,\dots,f_n$ be the candidate features and $y$ the binary label. A good panel contains features that are individually informative about $y$ and that are not informative about one another. We quantify both properties with mutual information. Each feature is discretized into $B=8$ equal-frequency bins, the mutual information $\MI(f_i;y)$ and $\MI(f_i;f_j)$ is estimated from the resulting histograms, and both quantities are scaled to $[0,1]$:
\begin{equation}
r_i = \frac{\MI(f_i;y)}{\max_{l}\MI(f_l;y)}, \qquad
R_{ij} = \frac{\MI(f_i;f_j)}{\max_{l<m}\MI(f_l;f_m)} \quad (i<j).
\end{equation}
Here $r_i$ is the relevance of feature $i$ and $R_{ij}$ the redundancy between features $i$ and $j$. A candidate panel is a binary vector $x\in\{0,1\}^n$, with $x_i=1$ if $f_i$ is selected. Following \citet{mucke2023}, its cost is
\begin{equation}\label{eq:qubo}
E_\alpha(x) = -\alpha\sum_{i=1}^{n} r_i x_i + (1-\alpha)\sum_{i<j} R_{ij}\, x_i x_j, \qquad \alpha\in[0,1].
\end{equation}
The first term rewards each selected feature by its relevance, and the second penalizes every selected pair by its redundancy. The cost is linear and quadratic in the binary variables and has no constraints, so minimizing it is a QUBO problem.

The parameter $\alpha$ sets the balance between the two terms and, indirectly, the size of the panel. When $\alpha$ is close to one, relevance dominates and adding features lowers the cost. When $\alpha$ is close to zero, redundancy dominates and small panels are preferred. Instead of fixing the panel size with a penalty term, which would add couplings between all pairs of variables, we search for a value of $\alpha$ by bisection until the global minimizer of Eq.~\eqref{eq:qubo} contains exactly $k$ features \citep{mucke2023}. For the instance sizes studied here, the global minimizer at each trial value of $\alpha$ is found by enumerating all $2^n$ configurations.

\subsection{Ising Hamiltonian}\label{subsec2.2}

To solve Eq.~\eqref{eq:qubo} on a quantum processor, each feature is assigned one qubit, and each candidate panel $x$ corresponds to the computational basis state $\ket{x}=\ket{x_1x_2\cdots x_n}$. A qubit in state $\ket{1}$ therefore means that the corresponding feature is selected. The cost function becomes a Hamiltonian $\Hp$ that is diagonal in this basis and satisfies $\Hp\ket{x}=E_\alpha(x)\ket{x}$, up to a constant energy shift. Its ground state is the optimal panel.

To build $\Hp$, we use the Pauli operator $Z_i$, which acts on qubit $i$ with eigenvalue $z_i=+1$ on $\ket{0}$ and $z_i=-1$ on $\ket{1}$. Each binary variable is then $x_i=(1-z_i)/2$. Substituting this relation into Eq.~\eqref{eq:qubo}, using $x_ix_j=(1-z_i-z_j+z_iz_j)/4$ and discarding constant terms gives the Ising Hamiltonian
\begin{equation}\label{eq:ising}
\Hp = \sum_{i} h_i Z_i + \sum_{i<j} J_{ij} Z_i Z_j,
\end{equation}
with local fields and couplings
\begin{equation}\label{eq:hJ}
h_i = \frac{\alpha r_i}{2} - \frac{1-\alpha}{4}\sum_{j\neq i} R_{ij}, \qquad
J_{ij} = \frac{(1-\alpha)R_{ij}}{4},
\end{equation}
where $R_{ij}=R_{ji}$. The relevance contribution to $h_i$ favours $z_i=-1$, which selects feature $i$. A positive $J_{ij}$ favours opposite spins. Together with the linear terms from the same substitution, it reproduces the redundancy penalty incurred when both features are selected. The discarded constant shifts every energy equally.

Finally, we rescale $\Hp$ so that the difference between its largest and smallest eigenvalues equals $n$. This normalization changes neither the eigenstates nor their ordering, and it keeps the useful range of the time step $\Delta t$ comparable across instances.

\subsection{Feedback-based quantum optimization}\label{subsec2.3}

Let $\ket{\psi_0}$ denote the state prepared at the start of a round. Standard FALQON uses a transverse-field driver $H_X=\sum_i X_i$ \citep{magann2022prl,magann2022pra}. For a fixed time step $\Delta t$, layer $\ell$ prepares the state
\begin{equation}\label{eq:layer}
 \ket{\psi_\ell}=e^{-i\beta_\ell H_X\Delta t}\,e^{-i\Hp\Delta t}\ket{\psi_{\ell-1}} .
\end{equation}
The rightmost operation applies a phase set by the energy of each candidate panel. The driver then rotates each qubit about its $X$ axis. Since $X_i$ connects panels that differ in the inclusion of feature $i$, these rotations can convert relative phases into changes in panel probabilities. The feedback signal measured after layer $\ell$ sets the amplitude of the next layer \citep{magann2022prl,magann2022pra},
\begin{equation}\label{eq:feedback}
 A_\ell=\bra{\psi_\ell}i[H_X,\Hp]\ket{\psi_\ell},\qquad \beta_{\ell+1}=-A_\ell .
\end{equation}
In continuous time, the feedback law gives $\mathrm{d}\langle\Hp\rangle/\mathrm{d}t=-A(t)^2\leq 0$. A finite time step can allow the energy to increase between successive circuit layers.

\subsection{Counterdiabatic-inspired drivers}\label{subsec2.4}

With only a few feedback layers per round, we ask whether the structure of the cost Hamiltonian can guide transitions between candidate panels. Counterdiabatic constructions suggest controls whose action on a feature depends on its local field and its couplings to other features. Consider the interpolation $H(\lambda)=(1-\lambda)H_X+\lambda\Hp$. Keeping the first term in the nested-commutator expansion of its adiabatic gauge potential gives \citep{claeys2019}
\begin{equation}\label{eq:agp}
 \mathcal A^{(1)}_\lambda=i\alpha_1(\lambda)[H(\lambda),\partial_\lambda H],
\end{equation}
where $\alpha_1(\lambda)$ is a scalar coefficient. The corresponding counterdiabatic term would be $\dot\lambda\mathcal A^{(1)}_\lambda$, with $\dot\lambda$ the rate of change of the interpolation parameter. Since $\partial_\lambda H=\Hp-H_X$ and $i[X_i,Z_i]=2Y_i$, the operator in Eq.~\eqref{eq:agp} has the exact expansion
\begin{equation}\label{eq:commutator}
\begin{aligned}
 i[H(\lambda),\partial_\lambda H]&=i[H_X,\Hp]\\
 &=2\sum_i h_iY_i+2\sum_{i<j}J_{ij}(Y_iZ_j+Z_iY_j).
\end{aligned}
\end{equation}
The first term weights each $Y$ rotation by the local Ising field. In the second term the rotation of one qubit depends on the $Z$ state of another. This dependence lets the driver distinguish the same feature flip in different candidate panels.

All four configurations retain the transverse-field driver $H_X$. Table~\ref{tab:controls} lists the additional control used in each case. Each additional driver $H_{\mathrm d}$ has a fixed gain $\gamma$ and its own layer amplitude. After layer $\ell$, we apply Eq.~\eqref{eq:feedback} separately to $H_X$ and to $H_{\mathrm d}$, substituting the latter for $H_X$ in the commutator to obtain the additional amplitude \citep{malla2024}. We use $h_{\max}=\max_i|h_i|$ to normalize the field-weighted operators.

\begin{table}[htbp]
\centering\small
\caption{Driver configurations.}\label{tab:controls}
\begin{tabular}{@{}lp{0.42\linewidth}p{0.28\linewidth}@{}}
\toprule
Configuration & Additional driver & Action on the features \\
\midrule
Transverse field & None & Single-bit flips from $H_X$ alone. \\[3pt]
Uniform $Y$ & $\gamma\sum_i Y_i$ & Equal weight on each qubit. \\[3pt]
Local $Y$ & $\frac{\gamma}{h_{\max}}\sum_i h_iY_i$ & Signed weights from the local fields. \\[3pt]
Two-body & $\frac{\gamma}{h_{\max}}\big[\sum_i h_iY_i$\newline $\quad +\sum_{i<j}J_{ij}(Y_iZ_j+Z_iY_j)\big]$ & Local rotations and rotations conditioned on another feature. \\
\bottomrule
\end{tabular}
\end{table}

For example, $Y_i Z_j$ acts as $+Y_i$ on qubit $i$ when feature $j$ is absent ($Z_j=+1$), and as $-Y_i$ when it is present ($Z_j=-1$). The coupling $J_{ij}$ sets the strength of this conditional contribution. The two-body variant retains these pair terms together with the local fields, as prescribed by the full commutator in Eq.~\eqref{eq:commutator}. Its operator structure is CD-inspired and its amplitude is determined by feedback.

The single-qubit terms commute within each one-body driver. For the two-body variant we group terms by the qubit carrying $Y_i$. Terms within each group commute and can be applied exactly. The ordered product of groups gives a first-order Trotter approximation to the driver evolution.

\subsection{Measurement-guided initialization}\label{subsec2.5}

\begin{figure}[tbp]
\centering
\includegraphics[width=\linewidth]{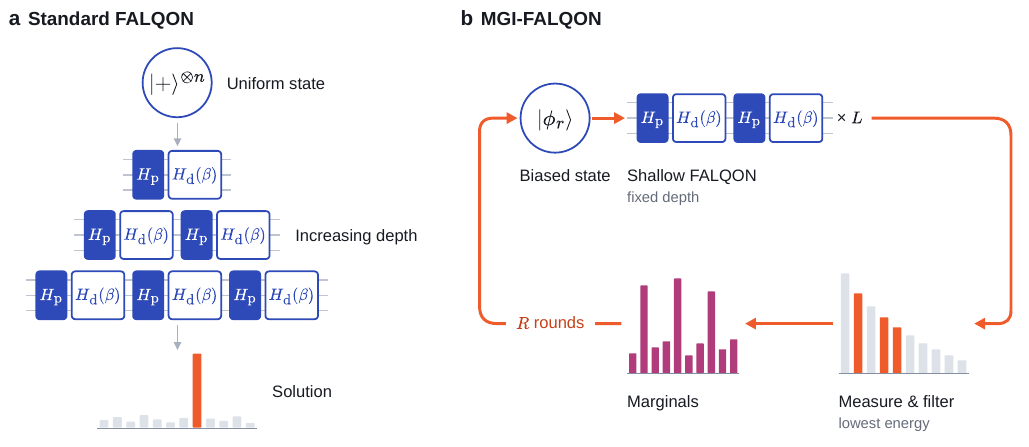}
\caption{Standard FALQON and measurement-guided initialization. (a) Standard FALQON starts in a uniform superposition. Each layer applies evolution under the problem Hamiltonian $H_{\mathrm p}$ and then under a driver $H_{\mathrm d}$ with amplitude $\beta_\ell$. Feedback measured after layer $\ell$ sets $\beta_{\ell+1}$ for the next layer as the circuit grows. (b) MGI-FALQON fixes the number of layers per circuit. At the end of each round, the inclusion frequencies of features among the retained low-energy outcomes, weighted by their probabilities, set the single-qubit biases of the next product state. The next round thus inherits information about promising features at the same circuit depth.}\label{fig:mgi}
\end{figure}

MGI starts the first round from $\ket{+}^{\otimes n}$, which assigns equal probability to all panels (Fig.~\ref{fig:mgi}). After the $L$th layer of round $r$, let $p_r(s)$ be the probability of bitstring $s$ in the computational basis. We first retain the most probable bitstrings above a probability threshold. We rank those candidates by QUBO energy and keep up to $n_r$ of the lowest-energy panels. The cutoff $n_r$ can decrease over successive rounds to narrow the set used for initialization. Denoting the resulting set by $\mathcal S_r$, we compute the filtered probability of including feature $i$ as
\begin{equation}\label{eq:marginals}
 c_i^{(r)}=\frac{\sum_{s\in\mathcal S_r}p_r(s)\,s_i}{\sum_{s\in\mathcal S_r}p_r(s)} .
\end{equation}
Here $s_i=1$ when panel $s$ includes feature $i$. Before the next round, qubit $i$ is prepared by applying $R_y(2\arcsin\sqrt{c_i^{(r)}})$ to $\ket{0}$. This product state carries forward the inclusion probability of each feature while discarding correlations between features \citep{rattighieri2026}. After the final round we use $p_R$ to evaluate bitstring recovery and the $k$ largest filtered marginals $c_i=c_i^{(R)}$ to select the classifier inputs.

If every filtered marginal is zero or one, the next round starts in a computational basis state. With the driver amplitudes initialized to zero, the cost evolution adds only a phase to that state. The feedback signals then vanish because the state is an eigenstate of $\Hp$, so subsequent rounds remain concentrated on the same bitstring.

\section{Experimental design}\label{sec3}

\subsection{Datasets and reference instances}\label{subsec3.1}

The UCI Cardiotocography (CTG) dataset contains 2,126 fetal heart-rate and uterine-contraction recordings with 21 derived variables and expert consensus labels \citep{ctg}. We classified 1,655 normal recordings against 471 labelled suspect or pathological. The autism metabolomics dataset comprises 83 participants with autism spectrum disorder and 76 neurotypical peers from the cohort of \citet{howsmon2017}. We used 22 biochemical measurements and ratios related to folate-dependent one-carbon metabolism and transsulfuration (FOCM/TS). We excluded the sibling group and the Vineland adaptive-behaviour score, which was collected only in the autism group. The unit of analysis was a recording for CTG and a participant for FOCM/TS.

We fixed the panel size at $k=4$ to compare selectors at a common level of compression, retaining approximately one fifth of the available variables in each dataset. The full-data instances served as solver benchmarks. For predictive evaluation, we constructed a new QUBO on each training fold, including its bin boundaries, mutual-information estimates and choice of $\alpha$. Exhaustive enumeration covered $2^{21}$ or $2^{22}$ bitstrings. This also makes the classical cost of choosing $\alpha$ part of the experimental procedure.

\subsection{State-vector simulations and readout}\label{subsec3.2}

We used Q-Tsuru \citep{qtsuru} on an NVIDIA GeForce RTX 3080 Ti. We initialized all driver amplitudes to zero at the start of every round, so the first layer applied only the cost evolution. The reference MGI runs used $R=15$ rounds of $L=5$ layers. In each round we retained up to 90 bitstrings with probability above $10^{-14}$, taking the most probable first. Of these, the energy filter kept up to $n_r=\lfloor 30-25(r-1)/(R-1)\rfloor$ candidates, decreasing from 30 in the first round to five in the last. For each driver we searched $\Delta t\in\{0.02,0.06,0.12,0.20,0.30\}$ and, when a CD-inspired driver was present, $\gamma\in\{0.5,1,2\}$. We minimized the final normalized energy gap, breaking ties in favour of the higher probability of the enumerated optimum. We repeated this search on the training portion of each fold in the five-fold MGI experiment. The depth scan used a separate search at each tested depth. To assess transfer of the full-data settings, we resampled records with replacement within each class and rebuilt the QUBO on each of 30 bootstrap samples per dataset. We kept the circuit parameters fixed throughout this comparison.

For final circuit probabilities $p_R(x)$, let $x^\star$ denote the bitstring returned by exact minimization. We recorded
\begin{equation}
\delta=\frac{\langle\Hp\rangle-E_{\min}}{E_{\max}-E_{\min}},\qquad
P_0=p_R(x^\star),\qquad P_4=\sum_{x:\,\sum_i x_i=4}p_R(x).
\end{equation}
For prediction, we took the four largest \emph{filtered} marginals $c_i$ after the last round. This always yields four variables, even when the most probable bitstring has a different cardinality. Panel agreement means that the four selected feature names form the same unordered set as the exact training-fold panel.

\subsection{Predictive evaluation and statistical comparisons}\label{subsec3.3}

In a stratified five-fold experiment, we compared the four MGI-FALQON configurations and standard QAOA with the exact training-fold panel. We included five-layer QAOA with a transverse-field mixer as a variational baseline for feature selection on the same QUBOs \citep{farhi2014}. Within MGI-FALQON, we compared each CD-inspired driver with the transverse field alone to assess its effect on optimal-panel recovery and held-out discrimination. For each selected panel we fitted standardization and logistic regression on the training data and measured AUC on the held-out fold.

The second experiment used ten repetitions of five-fold cross-validation to compare the exact QUBO with classical selectors: greedy minimum-redundancy maximum-relevance (mRMR), the four highest individual relevance scores, $\ell_1$-regularized logistic regression (LASSO) \citep{tibshirani1996}, and Shapley additive explanation (SHAP) rankings \citep{lundberg2017} derived from a random forest and from a logistic regression. All selectors returned four features. As references, we also evaluated a random four-feature panel, averaging its AUC over 100 draws within each split, and a logistic regression on all features. The quantum solvers were not run in this experiment. Its first repetition uses the splits of the five-fold experiment, so the exact QUBO and the other deterministic selectors have the same AUCs in both, and we treat the two summaries as overlapping samples.

For QAOA we used the same normalized Ising Hamiltonian as for MGI-FALQON. Starting from $\ket{+}^{\otimes n}$, layer $\ell$ applied $e^{-i\theta_\ell\Hp}$ followed by $e^{-i\eta_\ell H_X}$, and the $2p=10$ angles were chosen to minimize $\langle\Hp\rangle$. Like the MGI runs, QAOA was simulated with the full state vector and without shot noise. The energy was computed exactly from the final probabilities, and its gradient with respect to all angles was obtained by adjoint differentiation, which we checked against central finite differences on an eight-qubit test instance. The initial angles followed linear ramps inspired by Trotterized quantum annealing \citep{sack2021}, $\theta_\ell=s_\ell\tau$ and $\eta_\ell=\pm(1-s_\ell)\tau$ with $s_\ell=(\ell-1/2)/p$, for $\tau\in\{0.02,0.05,0.1,0.2,0.4,0.8\}$ and both signs of the mixer angle. The two ramps with the lowest energy and two random starts, drawn uniformly from $[-0.3,0.3]$ with a fixed seed, were each refined with the L-BFGS-B quasi-Newton method \citep{byrd1995} for at most 120 iterations, and the solution with the lowest energy was kept. The QAOA panel consisted of the four largest single-qubit marginals of the resulting distribution. Because QAOA has no energy filter, these marginals are unfiltered. The same procedure was applied to the full-data reference instances (Fig.~\ref{fig:qaoamarg}). The optimizer therefore had access to exact energies and gradients, a more favourable setting than on hardware, where both would have to be estimated from a finite number of shots.

We report arithmetic means and sample standard deviations across folds. Every comparison with the exact QUBO paired the AUCs obtained on the same split. Because the training sets of different folds overlap, we used the corrected resampled $t$ test \citep{nadeau2003}. Appendix~\ref{app:contrasts} describes the confidence intervals and adjustments for multiple comparisons.

\section{Results}\label{sec4}

\subsection{Recovery of the optimal feature panel}\label{subsec:solver}

\begin{figure}[H]
\centering
\includegraphics[width=0.96\linewidth]{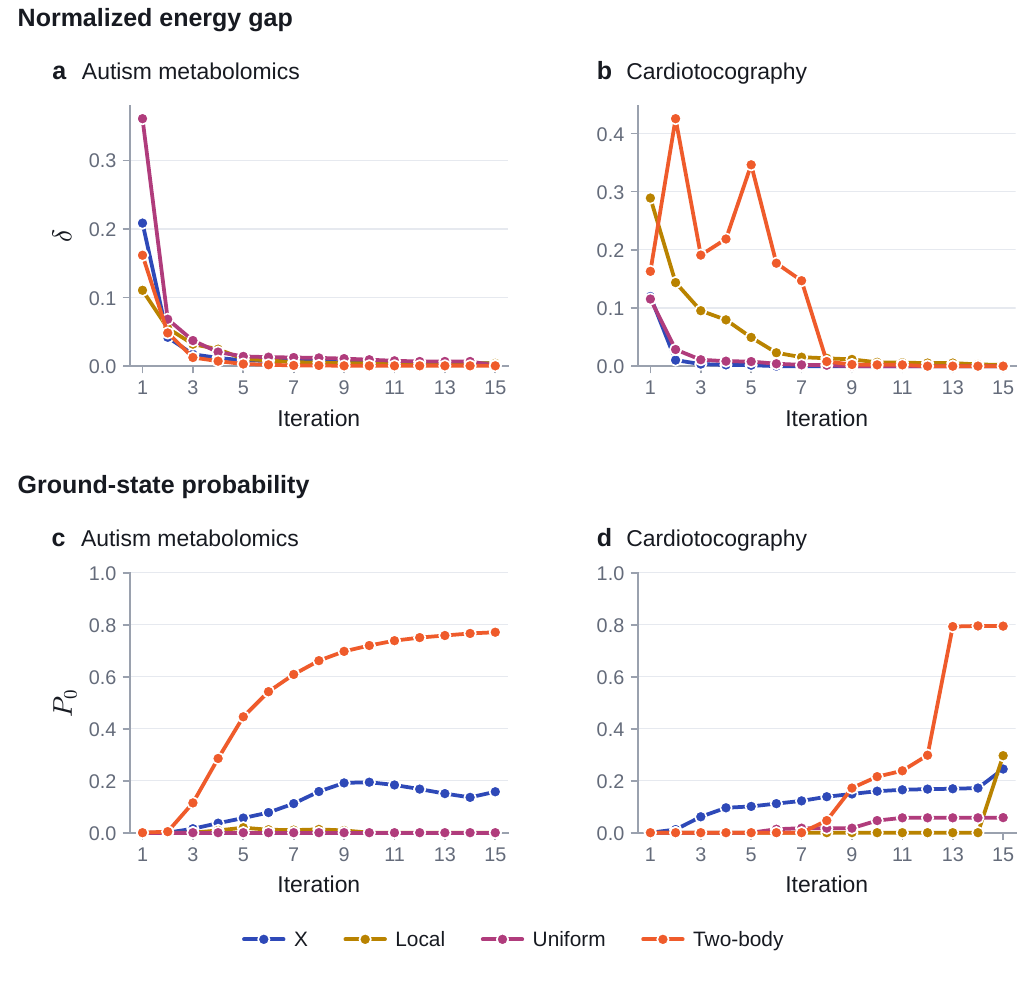}
\caption{Convergence over 15 rounds of five layers for the four drivers on FOCM/TS (left) and CTG (right). Panels (a,b) track the normalized mean-energy gap $\delta$ from the exact minimum. Panels (c,d) give the probability $P_0$ of sampling the exact optimal panel. Several drivers reach small gaps, yet the two-body driver assigns the largest final probability to the optimum on both reference instances. Curves were digitized from the simulation plots. Terminal probabilities reported in the text use tabulated outputs.}\label{fig:convergence}
\end{figure}

For the full CTG instance, bisection gave $\alpha=0.2188$ and selected accelerations (AC), severe decelerations (DS), prolonged decelerations (DP) and time with abnormal short-term variability (ASTV). For FOCM/TS, $\alpha=0.6250$ selected 8-OHG, total glutathione (tGSH), the free glutathione to oxidized glutathione ratio (fGSH/GSSG) and nitrotyrosine. These full-data solutions define the reference bitstrings in Fig.~\ref{fig:convergence}.

With five layers per round, the two-body driver raised $P_0$ from 0.2449 to 0.7949 on CTG and from 0.1575 to 0.7714 on FOCM/TS relative to the transverse field (Fig.~\ref{fig:convergence}). The ratio $P_0/P_4$ identifies where the gain occurred. On FOCM/TS, nearly every four-feature sample of the transverse-field circuit was already the optimum ($0.1575/0.1580$). The two-body driver primarily increased the chance of drawing four features at all. On CTG, the proportion of four-feature samples that matched the optimum also rose, from about 60\% with the transverse field to more than 99\% with the two-body driver (Table~\ref{tab:solver}). Thus an increase in sampling success can reflect different changes in the panel distribution.

The ranking reversed when the parameters chosen on the full datasets were transferred to resampled QUBOs. Across 30 class-stratified bootstrap samples per dataset, the transverse field had the higher mean $P_0$ (Table~\ref{tab:solver}). The bootstrap therefore measures how well a fixed setting transfers to a changed objective. It leaves open how the drivers would rank if retuned on each resample. All bootstrap instances had four-feature optima.

\begin{table}[!htbp]
\caption{MGI-FALQON performance on reference and bootstrap instances. Reference values after 15 five-layer rounds are rounded to four decimals. The final column averages $P_0$ over 30 class-stratified resamples. Bold marks the lowest $\delta$ and highest probabilities per dataset.}\label{tab:solver}
\centering\small
\setlength{\tabcolsep}{4pt}
\begin{tabular}{@{}lcccccc@{}}
\toprule
Driver & $\Delta t$ & $\gamma$ & $\delta$ & $P_0$ & $P_4$ & Bootstrap $P_0$ \\
\midrule
\multicolumn{7}{@{}l}{\textit{Cardiotocography}} \\
$X$       & 0.06 & --- & 0.0005 & 0.2449 & 0.4071 & \textbf{0.290} \\
Local $Y$ & 0.20 & 1   & \textbf{0.0000} & 0.2963 & 0.2963 & 0.141 \\
Uniform $Y$ & 0.30 & 1 & 0.0007 & 0.0580 & 0.1357 & 0.094 \\
Two-body & 0.30 & 1 & 0.0001 & \textbf{0.7949} & \textbf{0.7951} & 0.094 \\
\addlinespace
\multicolumn{7}{@{}l}{\textit{Autism metabolomics}} \\
$X$       & 0.02 & --- & 0.0004 & 0.1575 & 0.1580 & \textbf{0.317} \\
Local $Y$ & 0.20 & 1   & 0.0017 & 0.0000 & \textbf{0.9477} & 0.260 \\
Uniform $Y$ & 0.02 & 2 & 0.0012 & 0.0000 & 0.0000 & 0.145 \\
Two-body & 0.20 & 1 & \textbf{0.0003} & \textbf{0.7714} & 0.7714 & 0.084 \\
\bottomrule
\end{tabular}
\end{table}

\begin{figure}[H]
\centering\includegraphics[width=0.94\linewidth]{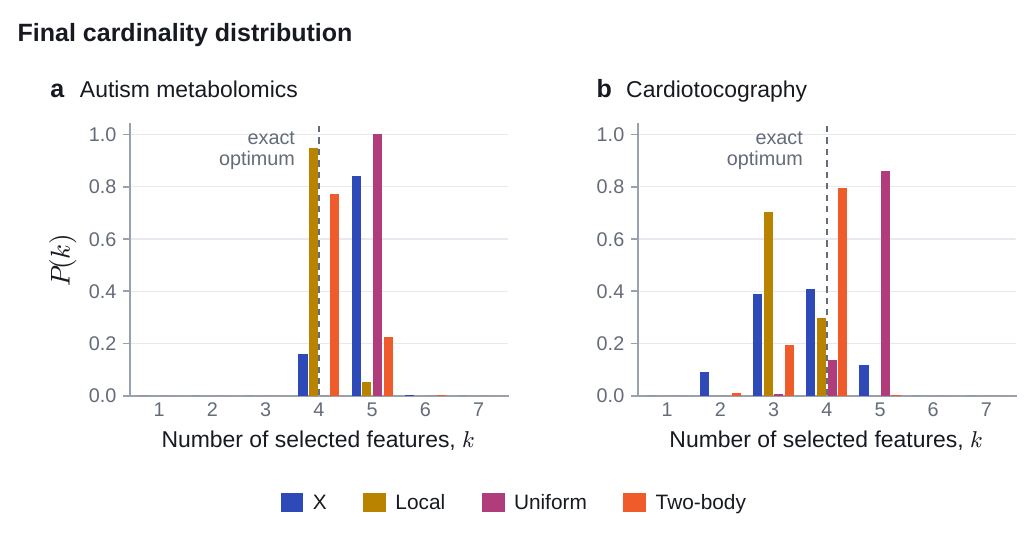}
\caption{Distribution of panel sizes after the final MGI round on (a) FOCM/TS and (b) CTG. For each driver, a bar at $k$ gives the probability of measuring a bitstring with $k$ selected features. The dashed line marks the target size of four. On FOCM/TS, the local $Y$ driver returns four features in about 95\% of outcomes, although almost none match the optimal quartet. The transverse-field bars for CTG were digitized.}\label{fig:cardinality}
\end{figure}

On FOCM/TS, the local $Y$ driver reached the target cardinality more often than the two-body driver (Fig.~\ref{fig:cardinality}). Its $P_0$ nevertheless rounded to zero, raising the question of which four-feature panels received that weight.

\begin{figure}[H]
\centering\includegraphics[width=0.94\linewidth]{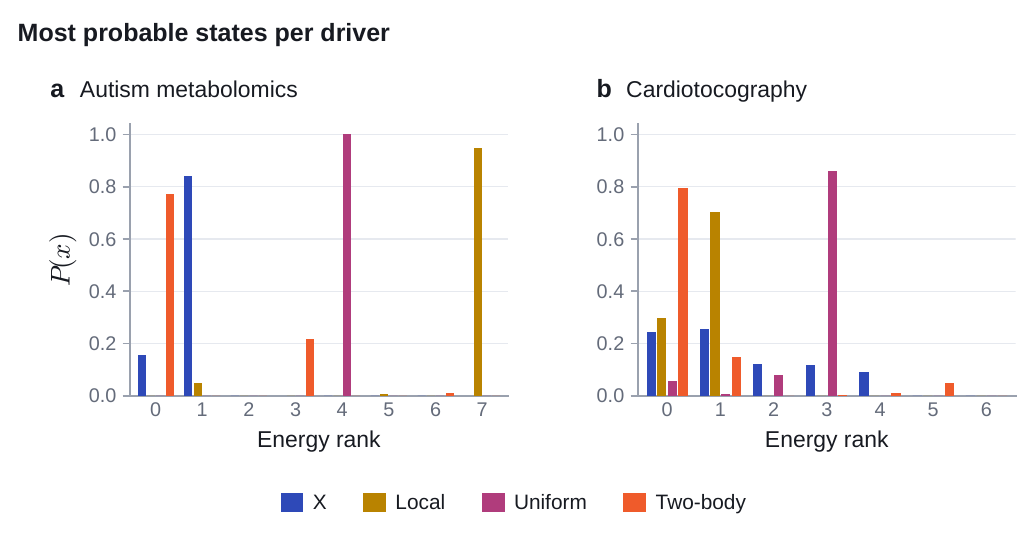}
\caption{Which panels carry the final measurement probability? For (a) FOCM/TS and (b) CTG, the five most probable bitstrings from each driver are pooled and ordered by tabulated QUBO energy, with ties broken by bitstring. Rank zero is the exact optimum. Bar heights give each driver's probability for the displayed panel. On FOCM/TS, the local $Y$ driver concentrates on a quartet that substitutes methionine for tGSH, explaining its high four-feature yield in Fig.~\ref{fig:cardinality}. The two-body driver concentrates on rank zero in both datasets.}\label{fig:topstates}
\end{figure}

The leading bitstrings resolve this discrepancy (Fig.~\ref{fig:topstates}). The local $Y$ driver assigned about 95\% of the probability to one panel that substituted methionine for tGSH. In this case the error involved the identity of one feature even though the circuit consistently selected the requested number.

\begin{figure}[H]
\centering\includegraphics[width=0.94\linewidth]{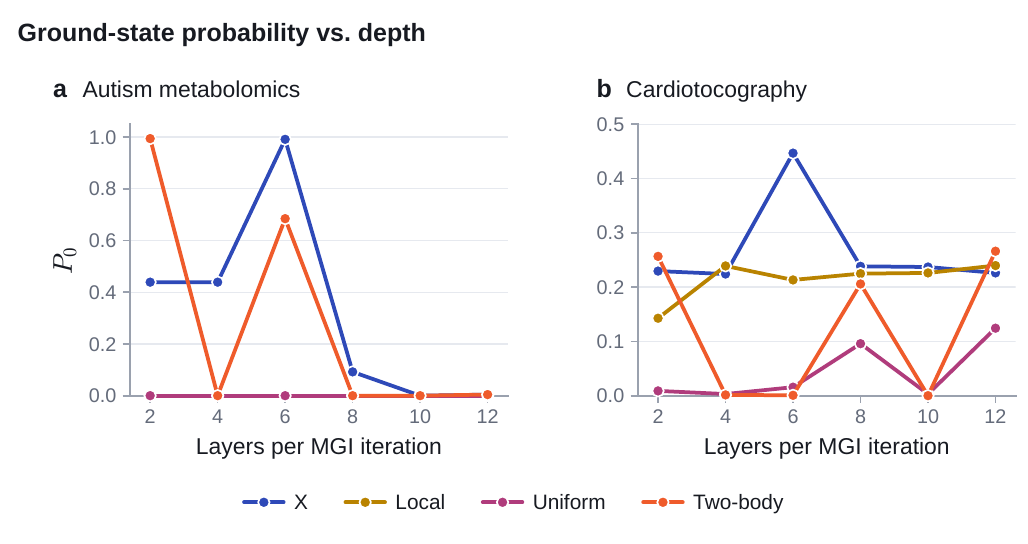}
\caption{Probability $P_0$ of sampling the optimal panel as the number of layers per MGI round varies on (a) FOCM/TS and (b) CTG. Symbols mark the six depths tested, with parameters tuned separately at each depth. Lines join the digitized values to guide the eye. On FOCM/TS, the transverse-field driver peaks near unity at six layers and falls below 0.10 at eight, illustrating the nonmonotonic dependence on depth.}\label{fig:depth}
\end{figure}

On FOCM/TS, the transverse-field driver reached $P_0\simeq0.99$ at six layers per round but fell below 0.10 at eight (Fig.~\ref{fig:depth}). Even with parameters tuned separately at each depth, adding layers did not consistently improve the probability of sampling the optimum.

\begin{figure}[H]
\centering\includegraphics[width=0.94\linewidth]{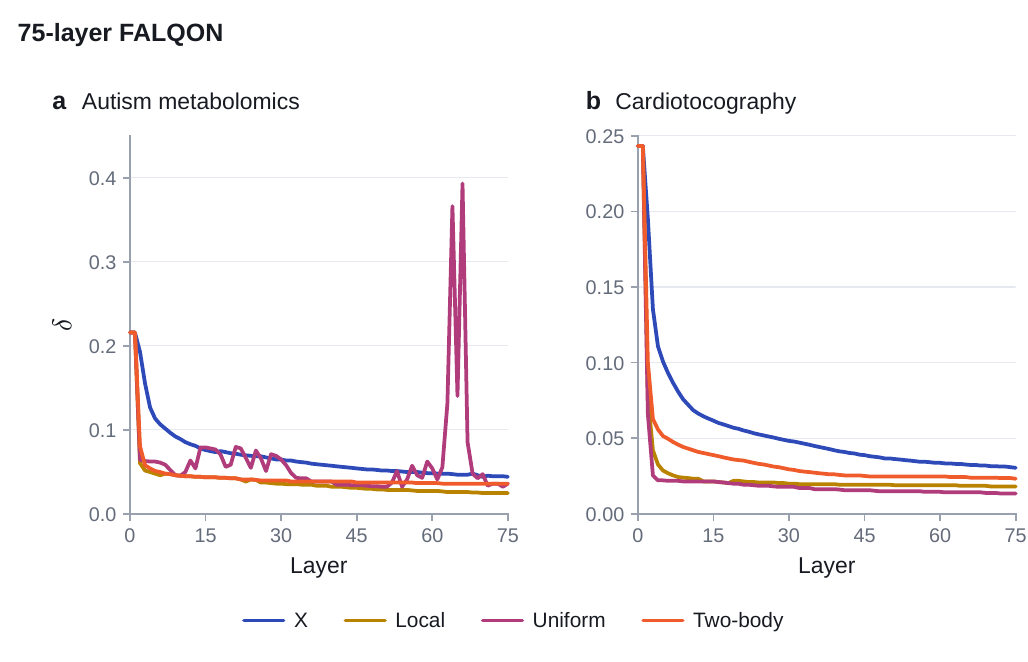}
\caption{Normalized mean-energy gap $\delta$ versus layer in a continuous 75-layer FALQON run on (a) FOCM/TS and (b) CTG. Each curve starts from the uniform product state and uses parameters tuned for its driver. The gap at layer 75 remains above the corresponding terminal gap for 15 five-layer MGI rounds in Table~\ref{tab:solver}. Both protocols schedule 75 layers in total, and only the MGI protocol filters outcomes and prepares a new state between rounds. Curves were digitized from the original trajectories.}\label{fig:long}
\end{figure}

Figure~\ref{fig:long} provides a second depth comparison. A continuous run of 75 feedback layers had a larger terminal normalized energy gap for every driver on both datasets than the corresponding 15 rounds of five layers (Table~\ref{tab:solver}). Both constructions schedule 75 layers in total. The repeated protocol also ranks candidates by classical energy and prepares 14 new product states, so the comparison measures the combined effect of filtering and restarting.

\FloatBarrier
\subsection{Predictive performance and panel stability}\label{subsec:panels}

Panel agreement and held-out AUC capture different consequences of a readout error (Table~\ref{tab:five}). The two-body readout recovered nine of ten exact training-fold panels. Its one mismatch, in CTG fold 3, exchanged severe decelerations (DS) for uterine contractions (UC), taking held-out AUC from 0.9416 to 0.9456. In the same fold QAOA also selected UC, but displaced abnormal short-term variability (ASTV) and gave an AUC of 0.891. Both are one-feature substitutions, yet they have different predictive consequences. On FOCM/TS three of the four MGI drivers returned the same five panels as exact enumeration and reproduced their fold-wise AUCs. None of the non-tied MGI comparisons with the exact QUBO reached the 0.05 level after Holm adjustment.

The QAOA baseline recovered the exact panel in two of five CTG folds and one of five FOCM/TS folds, fewer than any MGI driver. Its mean AUCs, 0.9238 on CTG and 0.9650 on FOCM/TS, were 0.0133 and 0.0102 below the exact QUBO. Neither difference was significant with the corrected paired $t$ test (unadjusted $p=0.41$ and $0.29$). In FOCM/TS, four of the five QAOA panels contained the ratio of total glutathione to oxidized glutathione (tGSH/GSSG), and the losses were smaller.

\begin{figure}[H]
\centering\includegraphics[width=0.94\linewidth]{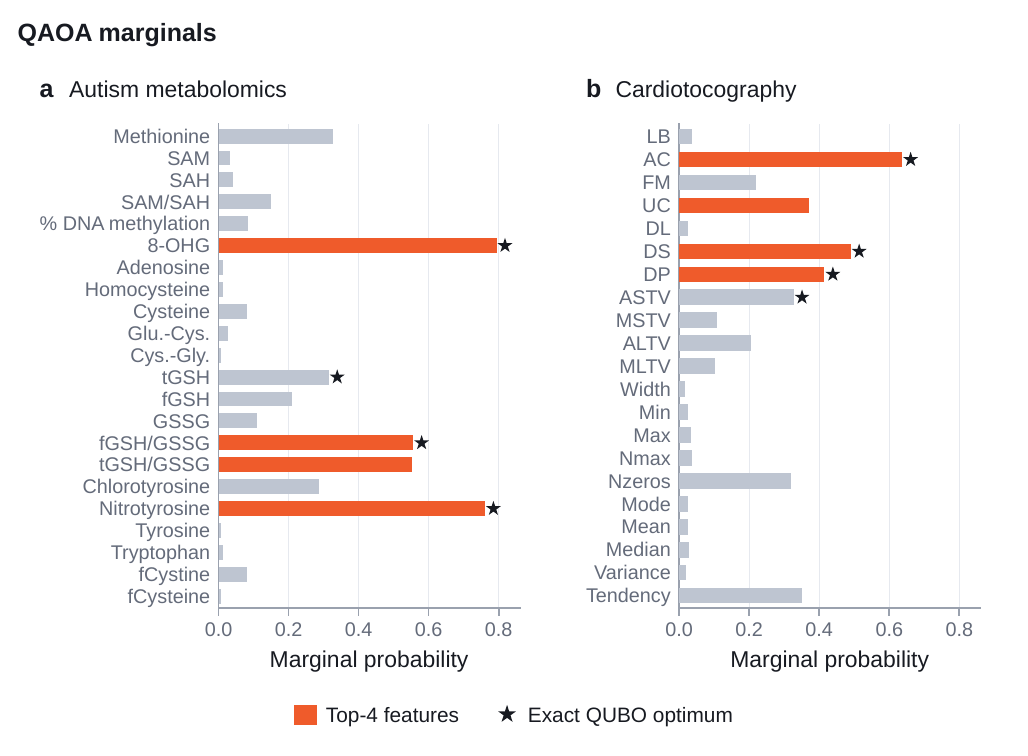}
\caption{Feature inclusion probabilities in five-layer QAOA on the full-data (a) FOCM/TS and (b) CTG instances. Each bar gives the probability that the corresponding feature bit is one. The four largest marginals (orange) define the selected panel, and stars identify features in the exact QUBO optimum. On CTG, UC ranks just above ASTV for the fourth slot, illustrating how a small difference between marginal probabilities can change the selected quartet.}\label{fig:qaoamarg}
\end{figure}

The full-data QAOA marginals illustrate how the readout selects the fourth feature on the reference instances (Fig.~\ref{fig:qaoamarg}). On CTG the inclusion probabilities for UC and ASTV were close (about 0.37 and 0.33), so a small change in their relative weights could change the fourth feature. On FOCM/TS, the readout selected the tGSH/GSSG ratio over tGSH itself. The two measurements share a biochemical component, but the ratio and the absolute concentration contribute differently to the relevance--redundancy objective.

For MGI-FALQON, the classifier uses the four largest marginals after energy filtering (Table~\ref{tab:five}). Appendix~\ref{app:marg} illustrates how the unfiltered circuit distribution can favour a different set of features.

\begin{table}[!htbp]
\caption{Panel recovery and held-out AUC across five folds. Each AUC entry is the mean $\pm$ sample SD of the held-out logistic-regression AUCs. Matches count folds recovering the exact four-feature set. QAOA uses five variational layers. Bold marks the best value in each column; tied values are all marked.}\label{tab:five}
\centering\small
\setlength{\tabcolsep}{4pt}
\begin{tabular}{@{}lcccc@{}}
\toprule
& \multicolumn{2}{c}{Cardiotocography} & \multicolumn{2}{c}{Autism metabolomics} \\
\cmidrule(lr){2-3}\cmidrule(l){4-5}
Selector & AUC & Matches & AUC & Matches \\
\midrule
Exact QUBO & $0.9370\,\pm\,0.0151$ & --- & \best{0.9752\,\pm\,0.0188} & --- \\
MGI, $X$ & $0.9371\,\pm\,0.0156$ & 3/5 & \best{0.9752\,\pm\,0.0188} & \textbf{5/5} \\
MGI, local $Y$ & $0.9371\,\pm\,0.0151$ & \textbf{5/5} & $0.9712\,\pm\,0.0150$ & 4/5 \\
MGI, uniform $Y$ & $0.9323\,\pm\,0.0240$ & 4/5 & \best{0.9752\,\pm\,0.0188} & \textbf{5/5} \\
MGI, two-body & \best{0.9379\,\pm\,0.0155} & 4/5 & \best{0.9752\,\pm\,0.0188} & \textbf{5/5} \\
QAOA, $p=5$ & $0.9238\,\pm\,0.0245$ & 2/5 & $0.9650\,\pm\,0.0186$ & 1/5 \\
\bottomrule
\end{tabular}
\end{table}

Beyond the five folds used to evaluate the quantum methods, we compared the exact QUBO with classical selectors across 50 training splits to assess panel stability and held-out discrimination. The exact QUBO selected AC, DP and ASTV in all 50 CTG training splits. The fourth feature alternated between DS (28 splits) and UC (22). On FOCM/TS, 19 distinct panels emerged. Nitrotyrosine appeared in 49 splits, 8-OHG in 41 and fGSH/GSSG in 37, yet the full-data quartet recurred only 15 times. The high mean held-out AUC on FOCM/TS (Table~\ref{tab:repeated}) therefore coexisted with considerable variation in the selected markers.

Using all measurements gave the highest mean AUC on both datasets. The mean gain over the exact four-feature panel was 0.0273 on CTG (95\% CI 0.0183 to 0.0364) and 0.0268 on FOCM/TS (95\% CI $-0.0004$ to 0.0541). The latter interval includes zero despite a similar point estimate.

\begin{table}[!htbp]
\caption{Held-out AUC across 50 splits. Values are logistic-regression means $\pm$ sample SD over ten repetitions of five-fold cross-validation. The random baseline averages 100 four-feature draws per split. All-feature models use 21 CTG or 22 FOCM/TS variables. Bold marks the highest four-feature mean per dataset.}\label{tab:repeated}
\centering\small
\begin{tabular}{@{}lcc@{}}
\toprule
Feature set or selector & Cardiotocography & Autism metabolomics \\
\midrule
Exact QUBO & $0.9364\,\pm\,0.0115$ & \best{0.9687\,\pm\,0.0271} \\
LASSO & \best{0.9440\,\pm\,0.0098} & $0.9593\,\pm\,0.0251$ \\
Greedy mRMR & $0.9130\,\pm\,0.0144$ & $0.9607\,\pm\,0.0261$ \\
Top-four relevance & $0.9130\,\pm\,0.0144$ & $0.9500\,\pm\,0.0341$ \\
SHAP, random forest & $0.9130\,\pm\,0.0144$ & $0.9537\,\pm\,0.0325$ \\
SHAP, logistic regression & $0.9322\,\pm\,0.0182$ & $0.9459\,\pm\,0.0340$ \\
Random four & $0.7901\,\pm\,0.0140$ & $0.9191\,\pm\,0.0192$ \\
\midrule
All features & $0.9637\,\pm\,0.0072$ & $0.9955\,\pm\,0.0056$ \\
\bottomrule
\end{tabular}
\end{table}

Among four-feature selectors, LASSO had the highest mean AUC on CTG. Its 0.0076 gain over the exact QUBO was sensitive to the family of comparisons used for Holm adjustment (Appendix~\ref{app:contrasts}). Greedy mRMR, relevance ranking and forest-based SHAP yielded identical fold-wise AUCs and lower means than the exact QUBO. On FOCM/TS, the exact QUBO ranked first by mean AUC, but only its advantage over random selection remained significant across the seven adjusted comparisons.

\section{Discussion}\label{sec5}

The choice of driver matters most when the required output is a sampled bitstring. The tuned two-body control increased the yield of exact solutions on both reference instances. A classifier, however, sees only the four features ranked highest after filtering. Once several circuits select the same measurements, increasing the probability of a particular bitstring offers little scope to improve their AUCs. The QAOA comparison illustrates the limits of attributing panel recovery to the feedback law alone. MGI used repeated preparation, a classical energy filter and a parameter grid, whereas QAOA optimized ten angles and used raw marginals. The comparison evaluates the two complete selection procedures at the stated settings.

Whether the QUBO itself is an appropriate target depends on the intended use of the panel. On CTG, exact optimization improved on the greedy mRMR filter, but logistic regression with LASSO selection reached a higher four-feature mean AUC. The QUBO scores associations of individual features with the label and penalizes pairwise redundancy. LASSO fits features jointly under a predictive loss, so the two objectives can prefer different subsets. Using all CTG features raised AUC further. On FOCM/TS, the exact quartet had the highest mean among compact panels, but the uncertainty was too large to resolve most differences between selectors. The repeated-split analysis benchmarks the exact objective and classical selectors. Direct evidence for the feedback readout comes from five training folds per dataset. Clinical transportability would require validation in independent cohorts.

Exact enumeration made it possible to calibrate $\alpha$ and evaluate recovery, while access to the state vector supplied exact feedback and identified the most probable candidates. A device would estimate both quantities from finite samples. Sampling can alter which candidates survive filtering, and those changes propagate into the next product state. The uniform and local $Y$ controls require only single-qubit rotations, whereas implementing the $Y_iZ_j$ and $Z_iY_j$ terms of the two-body driver can require additional entangling gates and routing on devices with limited connectivity. An equal number of feedback layers can therefore correspond to different circuit depths after compilation.

An informative control for a hardware study is to apply the same filter and reinitialization with identity evolution, using the same number of samples. This would isolate the contribution of feedback-driven evolution beyond repeated classical selection. The present simulations do not establish computational speedup.

\section{Conclusion}\label{sec6}

These simulations provide a proof of concept for using measurement-guided feedback to select compact panels of clinical variables. Its practical use here was to recover four-feature sets that could be evaluated directly on held-out observations. At five layers per round the two-body control increased the probability of sampling the optimum on both tuned reference instances. Its filtered readout recovered nine of ten training-fold optima, while several simpler drivers gave similar held-out discrimination. 

Future work will implement the feedback measurements, conditional rotations and MGI updates on real quantum hardware. The first test is whether a finite measurement budget preserves the ordering of the leading candidates well enough to prepare the next round. Comparing the transverse and two-body controls at matched shot budgets and reporting compiled two-qubit depth, exact-panel recovery and held-out AUC would show whether the additional interactions justify their cost on a device.

\backmatter

\bmhead{Acknowledgements}
The authors acknowledge support from the S\~ao Paulo Research Foundation (FAPESP). P.M.P., F.F.F., and C.J.V.-B. received support under Thematic Grant No.~2025/15490-0 (QuantaNet: Academia-Industry Network for Testing and Developing Quantum Technologies) of the QuTIa Program. G.S.F. received support under Grant No.~2025/19585-6, F.F.F. under Grant No.~2024/00998-6 through the Center for Research and Innovation on Smart and Quantum Materials (CRISQuaM), a Research, Innovation and Dissemination Center (CEPID), and C.J.V.-B. under Thematic Grant No.~2022/00209-6 (Second Generation Quantum Technologies). P.M.P. and C.J.V.-B. acknowledge support from CENPES/Petrobras under Grant No.~2025/00615-2. F.F.F. and C.J.V.-B. acknowledge support from the National Council for Scientific and Technological Development (CNPq) under Grant No.~408884/2024-0 for F.F.F. through the National Institute of Science and Technology for Applied Quantum Computing, and Grant No.~302234/2026-8 for C.J.V.-B. F.F.F. also acknowledges partial support from the Office of Naval Research (ONR), Project No.~N62909-24-1-2012.

\section*{Declarations}
\noindent\textbf{Funding.} This work received support from FAPESP under Grants No.~2025/15490-0 (QuantaNet: Academia-Industry Network for Testing and Developing Quantum Technologies), 2025/19585-6, 2024/00998-6 (CEPID CRISQuaM), and 2022/00209-6 (Second Generation Quantum Technologies). Additional support came from CENPES/Petrobras (Grant No.~2025/00615-2), CNPq (Grants No.~408884/2024-0 and 302234/2026-8), and the Office of Naval Research (Project No.~N62909-24-1-2012).
\par\noindent\textbf{Competing interests.} The authors declare no competing interests.
\par\noindent\textbf{Ethics approval and consent.} This study is a secondary analysis of publicly available, de-identified records.
\par\noindent\textbf{Consent for publication.} Not applicable.
\par\noindent\textbf{Data availability.} The cardiotocography data are available from the UCI Machine Learning Repository \citep{ctg}. The metabolomics data are available in the S1 Dataset of \citet{howsmon2017}. The fold-level AUC results, SHAP analyses and paired comparisons are available from the corresponding authors upon request.
\par\noindent\textbf{Code availability.} The code used for the simulations and data analyses is available from the corresponding authors upon request.

\begin{appendices}
\renewcommand{\theHfigure}{appendix.\thesection.\arabic{figure}}
\renewcommand{\theHtable}{appendix.\thesection.\arabic{table}}

\section{Paired AUC comparisons}\label{app:contrasts}
\setcounter{table}{0}

Each row of Table~\ref{tab:contrasts} compares one selector with the exact four-feature QUBO on the same 50 splits. A positive difference favours the listed selector. With $n$ paired differences and sample standard deviation $s_d$, the corrected standard error of the mean difference is $s_d\sqrt{1/n+1/4}$, where $1/4$ is the test-to-training size ratio for five folds, and the test has $n-1$ degrees of freedom. In the five-fold experiment ($n=5$), Holm's adjustment was applied within each dataset to all comparisons with the exact QUBO, excluding those whose AUCs coincided in every fold. The QAOA comparison is reported without adjustment. In the repeated experiment ($n=50$), Holm's adjustment was applied separately to the seven comparisons with the exact QUBO in each dataset. The individual 95\% confidence intervals use the corrected standard error.

The choice of multiplicity family affects one comparison. In the wider family of all pairwise comparisons among seven methods (omitting the top-four relevance ranking), the adjusted $p$ for CTG LASSO versus exact QUBO changes from 0.0366 to 0.0915. The estimated difference and its individual confidence interval are identical in the two families.

\begin{table}[!ht]
\caption{Paired differences in held-out AUC across 50 splits. The intervals account for overlapping training sets. Holm adjustment applies to the $p$ values for the seven reference comparisons in each dataset.}\label{tab:contrasts}
\centering\small
\setlength{\tabcolsep}{4pt}
\begin{tabular}{@{}lccc@{}}
\toprule
Comparator & $\Delta$AUC & Individual 95\% CI & Holm $p$ \\
\midrule
\multicolumn{4}{@{}l}{\textit{Cardiotocography}} \\
LASSO & $+0.0076$ & $[+0.0014, +0.0139]$ & $0.0366$ \\
Greedy mRMR & $-0.0234$ & $[-0.0335, -0.0132]$ & $1.31\times 10^{-4}$ \\
Top-four relevance & $-0.0234$ & $[-0.0335, -0.0132]$ & $1.31\times 10^{-4}$ \\
SHAP, random forest & $-0.0234$ & $[-0.0335, -0.0132]$ & $1.31\times 10^{-4}$ \\
SHAP, logistic regression & $-0.0042$ & $[-0.0208, +0.0125]$ & $0.6164$ \\
Random four & $-0.1463$ & $[-0.1585, -0.1341]$ & $5.71\times 10^{-28}$ \\
All features & $+0.0273$ & $[+0.0183, +0.0364]$ & $1.08\times 10^{-6}$ \\
\addlinespace
\multicolumn{4}{@{}l}{\textit{Autism metabolomics}} \\
LASSO & $-0.0094$ & $[-0.0410, +0.0222]$ & $1.000$ \\
Greedy mRMR & $-0.0080$ & $[-0.0235, +0.0075]$ & $1.000$ \\
Top-four relevance & $-0.0186$ & $[-0.0446, +0.0074]$ & $0.7843$ \\
SHAP, random forest & $-0.0149$ & $[-0.0416, +0.0117]$ & $1.000$ \\
SHAP, logistic regression & $-0.0227$ & $[-0.0683, +0.0229]$ & $1.000$ \\
Random four & $-0.0495$ & $[-0.0725, -0.0266]$ & $5.03\times 10^{-4}$ \\
All features & $+0.0268$ & $[-0.0004, +0.0541]$ & $0.3218$ \\
\bottomrule
\end{tabular}
\end{table}
\FloatBarrier

\section{Feature marginals before energy filtering}\label{app:marg}
\setcounter{figure}{0}
\renewcommand{\thefigure}{B\arabic{figure}}

\begin{figure}[H]
\centering\includegraphics[width=0.94\linewidth]{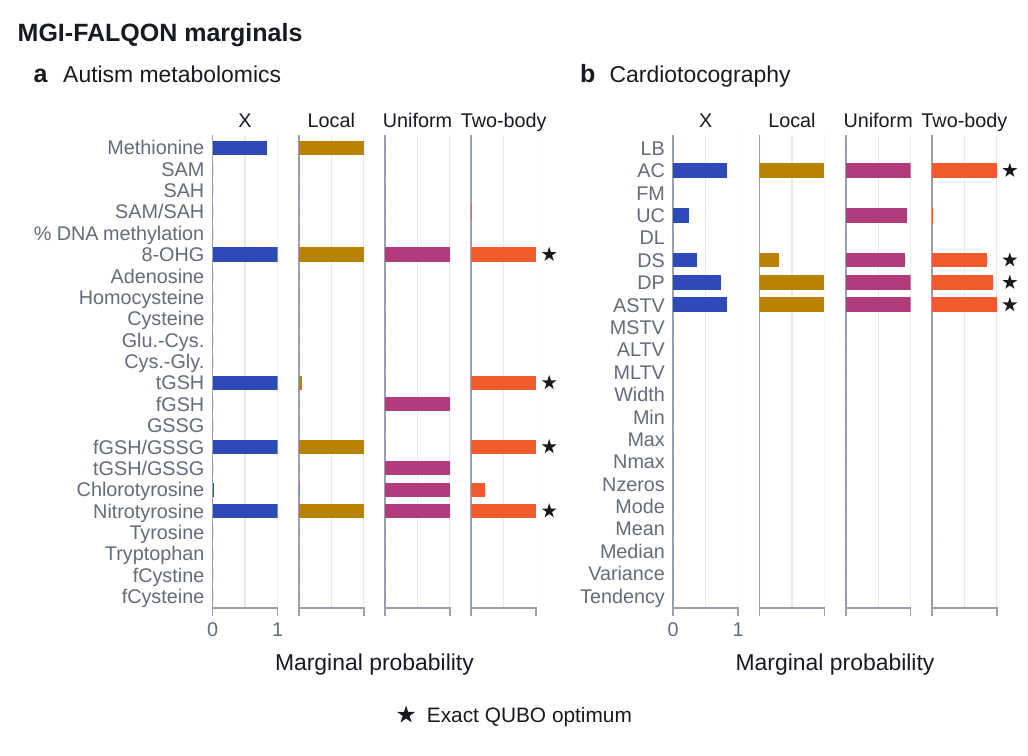}
\caption{Feature inclusion probabilities before energy filtering on (a) FOCM/TS and (b) CTG after 15 five-layer MGI rounds. The estimates use each driver's five most probable bitstrings, which carry nearly all the probability except for the CTG transverse field (83\%). Stars identify the exact-panel features. On FOCM/TS, methionine has a high transverse-field marginal because it occurs in the dominant five-feature bitstring. The classifier uses the four largest marginals after the energy filter in Eq.~\eqref{eq:marginals}.}\label{fig:mgimarg}
\end{figure}

Figure~\ref{fig:mgimarg} reconstructs feature inclusion probabilities from the leading bitstrings in Fig.~\ref{fig:topstates}. On FOCM/TS, the most probable transverse-field bitstring contains the four optimal features plus methionine. This explains the high marginal for methionine before filtering. Equation~\eqref{eq:marginals} gives the filtered probabilities used for panel selection.

\FloatBarrier

\end{appendices}

\makeatletter
\if@filesw\immediate\write\@auxout{\string\citation{bibsettings}}\fi
\makeatother
\bibliography{references}

@settings{bibsettings,
  key     = {0},
  options = {sort}
}

@article{guyon2003,
  author  = {Guyon, Isabelle and Elisseeff, Andr{\'e}},
  title   = {An introduction to variable and feature selection},
  journal = {Journal of Machine Learning Research},
  volume  = {3},
  pages   = {1157--1182},
  year    = {2003}
}

@article{saeys2007,
  author  = {Saeys, Yvan and Inza, I{\~n}aki and Larra{\~n}aga, Pedro},
  title   = {A review of feature selection techniques in bioinformatics},
  journal = {Bioinformatics},
  volume  = {23},
  number  = {19},
  pages   = {2507--2517},
  year    = {2007},
  doi     = {10.1093/bioinformatics/btm344}
}

@article{peng2005,
  author  = {Peng, Hanchuan and Long, Fuhui and Ding, Chris},
  title   = {Feature selection based on mutual information: criteria of max-dependency, max-relevance, and min-redundancy},
  journal = {IEEE Transactions on Pattern Analysis and Machine Intelligence},
  volume  = {27},
  number  = {8},
  pages   = {1226--1238},
  year    = {2005},
  doi     = {10.1109/TPAMI.2005.159}
}

@article{tibshirani1996,
  author  = {Tibshirani, Robert},
  title   = {Regression shrinkage and selection via the lasso},
  journal = {Journal of the Royal Statistical Society: Series B (Methodological)},
  volume  = {58},
  number  = {1},
  pages   = {267--288},
  year    = {1996},
  doi     = {10.1111/j.2517-6161.1996.tb02080.x}
}

@inproceedings{lundberg2017,
  author    = {Lundberg, Scott M. and Lee, Su-In},
  title     = {A unified approach to interpreting model predictions},
  booktitle = {Advances in Neural Information Processing Systems},
  volume    = {30},
  year      = {2017}
}

@article{lucas2014,
  author  = {Lucas, Andrew},
  title   = {Ising formulations of many {NP} problems},
  journal = {Frontiers in Physics},
  volume  = {2},
  pages   = {5},
  year    = {2014},
  doi     = {10.3389/fphy.2014.00005}
}

@article{glover2019,
  author  = {Glover, Fred and Kochenberger, Gary and Du, Yu},
  title   = {Quantum {Bridge} {Analytics} {I}: a tutorial on formulating and using {QUBO} models},
  journal = {4OR},
  volume  = {17},
  number  = {4},
  pages   = {335--371},
  year    = {2019},
  doi     = {10.1007/s10288-019-00424-y}
}

@article{mucke2023,
  author  = {M{\"u}cke, Sascha and Heese, Raoul and M{\"u}ller, Sabine and Wolter, Moritz and Piatkowski, Nico},
  title   = {Feature selection on quantum computers},
  journal = {Quantum Machine Intelligence},
  volume  = {5},
  number  = {1},
  pages   = {11},
  year    = {2023},
  doi     = {10.1007/s42484-023-00099-z}
}

@article{hellstern2024,
  author  = {Hellstern, Gerhard and Dehn, Vanessa and Zaefferer, Martin},
  title   = {Quantum computer based feature selection in machine learning},
  journal = {IET Quantum Communication},
  volume  = {5},
  number  = {3},
  pages   = {232--252},
  year    = {2024},
  doi     = {10.1049/qtc2.12086}
}

@article{felefly2023,
  author  = {Felefly, Tony and Roukoz, Camille and Fares, Georges and Achkar, Samir and Yazbeck, Sandrine and Meyer, Philippe and Kordahi, Manal and Azoury, Fares and Nasr, Dolly Nehme and Nasr, Elie and No{\"e}l, Georges and Francis, Ziad},
  title   = {An explainable {MRI}-radiomic quantum neural network to differentiate between large brain metastases and high-grade glioma using quantum annealing for feature selection},
  journal = {Journal of Digital Imaging},
  volume  = {36},
  number  = {6},
  pages   = {2335--2346},
  year    = {2023},
  doi     = {10.1007/s10278-023-00886-x}
}

@article{nau2025,
  author  = {Nau, Merlin A. and Nutricati, Luca A. and Camino, Bruno and Warburton, Paul A. and Maier, Andreas K.},
  title   = {Quantum annealing feature selection on light-weight medical image datasets},
  journal = {Scientific Reports},
  volume  = {15},
  pages   = {28937},
  year    = {2025},
  doi     = {10.1038/s41598-025-14611-x}
}

@article{orquin2026,
  author = {Orqu\'in-Marqu\'es, Jos\'e J. and Flores-Garrig\'os, Carlos and Gomez Cadavid, Alejandro and Simen, Anton and Solano, Enrique and Hegade, Narendra N. and Mart\'in-Guerrero, Jos\'e D. and Vives-Gilabert, Yolanda},
  title = {Analog quantum feature selection with neutral-atom quantum processors},
  journal = {Quantum Machine Intelligence},
  year = {2026},
  volume = {8},
  pages = {108},
  doi = {10.1007/s42484-026-00456-8}
}

@misc{liu2026,
  author        = {Liu, Hongbin and Lahmann, Robert and Campbell, Benjamin and Zhang, Zhemin and Liang, Zhiding and Bapat, Siona and Hahn, Juergen},
  title         = {Quantum feature selection for biomedical data analysis},
  year          = {2026},
  eprint        = {2609.30256},
  archivePrefix = {arXiv},
  primaryClass  = {quant-ph},
  howpublished  = {\href{https://arxiv.org/abs/2609.30256}{arXiv:2609.30256}}
}

@misc{farhi2014,
  author        = {Farhi, Edward and Goldstone, Jeffrey and Gutmann, Sam},
  title         = {A quantum approximate optimization algorithm},
  year          = {2014},
  eprint        = {1411.4028},
  archivePrefix = {arXiv},
  primaryClass  = {quant-ph},
  howpublished  = {\href{https://arxiv.org/abs/1411.4028}{arXiv:1411.4028}}
}

@article{cerezo2021,
  author  = {Cerezo, M. and Arrasmith, Andrew and Babbush, Ryan and Benjamin, Simon C. and Endo, Suguru and Fujii, Keisuke and McClean, Jarrod R. and Mitarai, Kosuke and Yuan, Xiao and Cincio, Lukasz and Coles, Patrick J.},
  title   = {Variational quantum algorithms},
  journal = {Nature Reviews Physics},
  volume  = {3},
  number  = {9},
  pages   = {625--644},
  year    = {2021},
  doi     = {10.1038/s42254-021-00348-9}
}

@article{mcclean2018,
  author  = {McClean, Jarrod R. and Boixo, Sergio and Smelyanskiy, Vadim N. and Babbush, Ryan and Neven, Hartmut},
  title   = {Barren plateaus in quantum neural network training landscapes},
  journal = {Nature Communications},
  volume  = {9},
  pages   = {4812},
  year    = {2018},
  doi     = {10.1038/s41467-018-07090-4}
}

@article{bittel2021,
  author  = {Bittel, Lennart and Kliesch, Martin},
  title   = {Training variational quantum algorithms is {NP}-hard},
  journal = {Physical Review Letters},
  volume  = {127},
  number  = {12},
  pages   = {120502},
  year    = {2021},
  doi     = {10.1103/PhysRevLett.127.120502}
}

@article{magann2022prl,
  author  = {Magann, Alicia B. and Rudinger, Kenneth M. and Grace, Matthew D. and Sarovar, Mohan},
  title   = {Feedback-based quantum optimization},
  journal = {Physical Review Letters},
  volume  = {129},
  number  = {25},
  pages   = {250502},
  year    = {2022},
  doi     = {10.1103/PhysRevLett.129.250502}
}

@article{magann2022pra,
  author  = {Magann, Alicia B. and Rudinger, Kenneth M. and Grace, Matthew D. and Sarovar, Mohan},
  title   = {Lyapunov-control-inspired strategies for quantum combinatorial optimization},
  journal = {Physical Review A},
  volume  = {106},
  number  = {6},
  pages   = {062414},
  year    = {2022},
  doi     = {10.1103/PhysRevA.106.062414}
}

@misc{rattighieri2026,
  shortauthor   = {Rattighieri, Prado et~al.},
  author        = {Rattighieri, Lucas A. M. and Prado, Pedro M. and {de Oliveira}, Marcos C. and Fanchini, Felipe F.},
  title         = {Measurement-guided state refinement for shallow feedback-based quantum optimization algorithm},
  year          = {2026},
  eprint        = {2602.20407},
  archivePrefix = {arXiv},
  primaryClass  = {quant-ph},
  howpublished  = {\href{https://arxiv.org/abs/2602.20407}{arXiv:2602.20407}}
}

@article{demirplak2003,
  author  = {Demirplak, Mustafa and Rice, Stuart A.},
  title   = {Adiabatic population transfer with control fields},
  journal = {The Journal of Physical Chemistry A},
  volume  = {107},
  number  = {46},
  pages   = {9937--9945},
  year    = {2003},
  doi     = {10.1021/jp030708a}
}

@article{berry2009,
  author  = {Berry, M. V.},
  title   = {Transitionless quantum driving},
  journal = {Journal of Physics A: Mathematical and Theoretical},
  volume  = {42},
  number  = {36},
  pages   = {365303},
  year    = {2009},
  doi     = {10.1088/1751-8113/42/36/365303}
}

@article{sels2017,
  author  = {Sels, Dries and Polkovnikov, Anatoli},
  title   = {Minimizing irreversible losses in quantum systems by local counterdiabatic driving},
  journal = {Proceedings of the National Academy of Sciences},
  volume  = {114},
  number  = {20},
  year    = {2017},
  doi     = {10.1073/pnas.1619826114}
}

@article{claeys2019,
  author  = {Claeys, Pieter W. and Pandey, Mohit and Sels, Dries and Polkovnikov, Anatoli},
  title   = {Floquet-engineering counterdiabatic protocols in quantum many-body systems},
  journal = {Physical Review Letters},
  volume  = {123},
  number  = {9},
  pages   = {090602},
  year    = {2019},
  doi     = {10.1103/PhysRevLett.123.090602}
}

@article{hegade2021,
  author  = {Hegade, Narendra N. and Paul, Koushik and Ding, Yongcheng and Sanz, Mikel and Albarr{\'a}n-Arriagada, F. and Solano, Enrique and Chen, Xi},
  title   = {Shortcuts to adiabaticity in digitized adiabatic quantum computing},
  journal = {Physical Review Applied},
  volume  = {15},
  number  = {2},
  pages   = {024038},
  year    = {2021},
  doi     = {10.1103/PhysRevApplied.15.024038}
}

@article{chandarana2022,
  author  = {Chandarana, P. and Hegade, N. N. and Paul, K. and Albarr{\'a}n-Arriagada, F. and Solano, E. and del Campo, A. and Chen, Xi},
  title   = {Digitized-counterdiabatic quantum approximate optimization algorithm},
  journal = {Physical Review Research},
  volume  = {4},
  number  = {1},
  pages   = {013141},
  year    = {2022},
  doi     = {10.1103/PhysRevResearch.4.013141}
}

@article{cadavid2025,
  author  = {Gomez Cadavid, Alejandro and Dalal, Archismita and Simen, Anton and Solano, Enrique and Hegade, Narendra N.},
  title   = {Bias-field digitized counterdiabatic quantum optimization},
  journal = {Physical Review Research},
  volume  = {7},
  year    = {2025},
  doi     = {10.1103/PhysRevResearch.7.L022010}
}

@article{malla2024,
  author  = {Malla, Rajesh K. and Sukeno, Hiroki and Yu, Hongye and Wei, Tzu-Chieh and Weichselbaum, Andreas and Konik, Robert M.},
  title   = {Feedback-based quantum algorithm inspired by counterdiabatic driving},
  journal = {Physical Review Research},
  volume  = {6},
  number  = {4},
  pages   = {043068},
  year    = {2024},
  doi     = {10.1103/PhysRevResearch.6.043068}
}

@article{gupta2025,
  author  = {Gupta, Riddhi S. and Wood, Carolyn E. and Engstrom, Teyl and Pole, Jason D. and Shrapnel, Sally},
  title   = {A systematic review of quantum machine learning for digital health},
  journal = {npj Digital Medicine},
  volume  = {8},
  pages   = {237},
  year    = {2025},
  doi     = {10.1038/s41746-025-01597-z}
}

@article{schuld2022,
  author  = {Schuld, Maria and Killoran, Nathan},
  title   = {Is quantum advantage the right goal for quantum machine learning?},
  journal = {PRX Quantum},
  volume  = {3},
  number  = {3},
  pages   = {030101},
  year    = {2022},
  doi     = {10.1103/PRXQuantum.3.030101}
}

@misc{bowles2024,
  author        = {Bowles, Joseph and Ahmed, Shahnawaz and Schuld, Maria},
  title         = {Better than classical? {The} subtle art of benchmarking quantum machine learning models},
  year          = {2024},
  eprint        = {2403.07059},
  archivePrefix = {arXiv},
  primaryClass  = {quant-ph},
  howpublished  = {\href{https://arxiv.org/abs/2403.07059}{arXiv:2403.07059}}
}

@misc{qtsuru,
  author       = {Rattighieri, Lucas},
  title        = {{Q-Tsuru}: quantum circuit and algorithm simulation library using {PyTorch}},
  year         = {2026},
  howpublished = {GitHub repository, version 1.0.0, \url{https://github.com/Lucas-Rattighieri/qtsuru}}
}

@misc{ctg,
  author = {Campos, D. and Bernardes, J.},
  title = {Cardiotocography},
  year = {2000},
  howpublished = {UCI Machine Learning Repository},
  doi = {10.24432/C51S4N}
}

@article{howsmon2017,
  author = {Howsmon, Daniel P. and Kruger, Uwe and Melnyk, Stepan and James, S. Jill and Hahn, Juergen},
  title = {Classification and adaptive behavior prediction of children with autism spectrum disorder based upon multivariate data analysis of markers of oxidative stress and {DNA} methylation},
  journal = {PLoS Computational Biology},
  year = {2017},
  volume = {13},
  number = {3},
  pages = {e1005385},
  doi = {10.1371/journal.pcbi.1005385}
}

@article{nadeau2003,
  author = {Nadeau, Claude and Bengio, Yoshua},
  title = {Inference for the generalization error},
  journal = {Machine Learning},
  volume = {52},
  pages = {239--281},
  year = {2003},
  doi = {10.1023/A:1024068626366}
}

@article{prado2026,
  author  = {Prado, Pedro M. and Rattighieri, Lucas A. M. and {do Carmo}, Rafael Sim{\~o}es and Franco, Giovanni S. and Pexe, Guilherme E. L. and Drinko, Alexandre and Dorlass, Erick G. and {de Almeida}, Tatiana F. and Fanchini, Felipe F.},
  title   = {Quantum feedback algorithms for {DNA} assembly using {FALQON} variants},
  journal = {Quantum Machine Intelligence},
  year    = {2026},
  note    = {\href{https://arxiv.org/abs/2602.21080}{arXiv:2602.21080}}
}

@article{prado2026qip,
  author  = {Prado, Pedro M. and {do Carmo}, Rafael Sim{\~o}es and Rattighieri, Lucas A. M. and Arruda, Luiz Gustavo Esmenard and Franco, Giovanni S. and {de Oliveira}, Marcos C{\'e}sar and Fanchini, Felipe F.},
  title   = {Portfolio optimization via {FALQON} variants: warm starting, {XY} mixing, and circuit depth reduction},
  journal = {Quantum Information Processing},
  volume  = {25},
  number  = {8},
  pages   = {275},
  year    = {2026},
  doi     = {10.1007/s11128-026-05291-1}
}

@article{boev2021,
  author  = {Boev, A. S. and Rakitko, A. S. and Usmanov, S. R. and Kobzeva, A. N. and Popov, I. V. and Ilinsky, V. V. and Kiktenko, E. O. and Fedorov, A. K.},
  title   = {Genome assembly using quantum and quantum-inspired annealing},
  journal = {Scientific Reports},
  volume  = {11},
  pages   = {13183},
  year    = {2021},
  doi     = {10.1038/s41598-021-88321-5}
}

@article{sarkar2021,
  author  = {Sarkar, Aritra and Al-Ars, Zaid and Bertels, Koen},
  title   = {{QuASeR}: Quantum accelerated de novo {DNA} sequence reconstruction},
  journal = {PLoS ONE},
  volume  = {16},
  number  = {4},
  pages   = {e0249850},
  year    = {2021},
  doi     = {10.1371/journal.pone.0249850}
}

@article{robert2021,
  author  = {Robert, Anton and Barkoutsos, Panagiotis Kl. and Woerner, Stefan and Tavernelli, Ivano},
  title   = {Resource-efficient quantum algorithm for protein folding},
  journal = {npj Quantum Information},
  volume  = {7},
  pages   = {38},
  year    = {2021},
  doi     = {10.1038/s41534-021-00368-4}
}

@article{sack2021,
  author  = {Sack, Stefan H. and Serbyn, Maksym},
  title   = {Quantum annealing initialization of the quantum approximate optimization algorithm},
  journal = {Quantum},
  volume  = {5},
  pages   = {491},
  year    = {2021},
  doi     = {10.22331/q-2021-07-01-491}
}

@article{byrd1995,
  author  = {Byrd, Richard H. and Lu, Peihuang and Nocedal, Jorge and Zhu, Ciyou},
  title   = {A limited memory algorithm for bound constrained optimization},
  journal = {SIAM Journal on Scientific Computing},
  volume  = {16},
  number  = {5},
  pages   = {1190--1208},
  year    = {1995},
  doi     = {10.1137/0916069}
}
\end{document}